\documentclass[11pt]{article}
\usepackage{vietnam,epsfig}

\def\Journal#1#2#3#4{{#1} {\bf #2}, #3 (#4)}

\def\be{\begin{equation}}
\def\ee{\end{equation}}
\def\bea{\begin{eqnarray}}
\def\eea{\end{eqnarray}}

\begin{document}
\vspace*{4cm}
\title{DIRECT $CP$ VIOLATION - RECENT RESULTS FROM BABAR}

\author{ A. SATPATHY }

\address{Department of Physics, University of Texas at Austin, Texas  78712-0264, USA \\
 ( For the BaBar Collaboration )}

\maketitle\abstracts{
Measurements of the CKM parameter $\sin(2 \beta)$ have established $CP$ violation in the $B^0$ meson
system arising from the interference between mixing and decay. However, direct 
$CP$ violation, arising from the interference among different
terms in the decay amplitude, had not been observed so far. 
We report a first observation of direct $CP$ violation in $B^{0} \to K^{+} \pi^{-}$ decays with the
BaBar detector. Other selected results based on the search for direct $CP$ violation in 
several other B decays are also presented.}

\section{Introduction}
We use the term ``direct $CP$ Violation'' for
$CP$ violation in meson decays, when the $CP$ violation appears as a result of interference among
various terms in decay amplitude and will not occur unless at least two terms
have different weak phases and different strong phases. For a decay process, $B \to f$, and its
charge conjugate $\bar{B} \to \bar{f}$, the
direct $CP$ asymmetry is defined by
\begin{equation}
A_{CP} = \frac{\Gamma (\bar{B} \to \bar{f}) - \Gamma (B \to f)}{\Gamma (\bar{B} \to \bar{f}) + \Gamma (B \to f)}
\label{eq:acp1}
\end{equation}
Here $B$ refers to either a charged or a neutral $B$ meson. With
the decay amplitudes $A_{f} = |A_{f}| e^{i \phi_{weak}} e^{i \theta_{strong}}$ and
$\bar{A}_{\bar{f}} = |A_{\bar{f}}| e^{-i \phi_{weak}} e^{i \theta_{strong}}$, 
and $R =  |A_{{f}}|/|\bar{A}_{\bar{f}}|$, Eq.~\ref{eq:acp1} becomes
\begin{equation}
A_{CP} = \left [ \frac{2}{R + 1/R + \cos(\Delta \phi_{weak}) \cos(\Delta \theta_{strong})}\right ] \!
\sin(\Delta \phi_{weak}) \sin(\Delta \theta_{strong})
\label{eq:acp2}
\end{equation}
Given a measurement of the $CP$ asymmetry, the interpretation of the result
depends on the relative strong phase ($\theta_{strong}$) that arises from the 
final state interactions which are a not well understood. 
It is therefore important to have a large variety of experimental inputs to 
better understand the non-perturbative physics leading to the occurrence of 
strong phase and to further confirm or refute the Kobayashi-Maskawa
picture of $CP$ violation in Standard Model(SM)~\cite{km}.

Depending on the model used, expected $CP$ asymmetries in $B$ 
meson decays vary widely.
An asymmetry as small as $2 - 10$\% is expected in a factorization model 
calculation~\cite{ali_benke}, while new physics could introduce new large
phases directly leading to an expected asymmetry of $40 - 60$\%~\cite{hou}.
In some classes of $B$ decays (such as radiative $B \to X_{s} \gamma$), 
the expected $CP$ asymmetry is less than 1\%~\cite{radiative}. 
A measurement of significant non-zero 
$CP$ asymmetry  will be an evidence for the contribution of new physics in such $B$ decays. 
The dedicated physics program at the
$B$ factories will continue to do stringent tests of the various models and 
improve our understanding of the source of $CP$ violation significantly.
In this article, we will briefly review some of the ongoing 
search for direct $CP$ violation in $B$ decays 
with the BaBar detector~\cite{babar} at the Stanford Linear Accelerator Center (SLAC)
PEP-II $e^+ e^-$ asymmetric-energy storage ring.

\section{General Analysis Procedure}
For each event, charged tracks and neutral particles in the detector are identified using various quality
requirements on reconstructed tracks and neutral showers. $B$ candidates are selected
using two kinematic variables $M_{ES} = \sqrt{E^{2}_{beam} - p^{*2}_{B}}$,
the energy substituted mass and $\Delta E = E_{beam} - E^{*}_{B}$, where $E^{*}_{beam}$ is
the beam energy and $p^{*}_{B} (E^{*}_{B})$ is the measured momentum (energy) of the $B$ candidate in 
the $\Upsilon(4S)$ center-of-mass(CM) frame. While $M_{ES}$ expresses the momentum
conservation in the decay, $\Delta E$ expresses the energy conservation of the particles
in the decay and is sensitive to the missing particles and $K/\pi$ misidentification.
In most of the cases considered here, the analyses are affected by the dominant
source of background arising from $e^{+} e^{-} \to q \bar{q}(q=u, d, c, s)$ transitions.
To reject this, we exploit the difference 
in topology between jetty hadronization of continuum events and spherical
decays of $B$'s on the $\Upsilon(4S)$ CM frame. The topology is described 
using the angle $\theta_{T}$ between the thrust axis of the $B$ candidate and the thrust axis
of the charged and neutral particles in the rest of the event (ROE)~\cite{babar}.
Sometimes the angle $\theta_{S}$, defined in the CM frame, between 
sphericity axis~\cite{sphericity} of the $B$
candidate and the sphericity axis of the ROE is also used to discriminate signal from continuum
background. For background events, $|\cos \theta_{S}|$ peaks sharply
near unity, while it is nearly uniform for signal events.
Other useful quantity that characterize the event topology are two sums over
the ROE: $L_{0} = \sum |\vec{p_{i}}^{*}|$ and $L_{2} = \sum |\vec{p_{i}}^{*}|\cos^{2} \theta_{i}$, 
where $\theta_{i}$ is the angle between the momentum $\vec{p_{i}}^{*}$ and the thrust axis
of the $B$ candidate. Additional separation is achieved using the angle $\theta_{B}$ between the $B$
momentum direction and the beam axis. To maximize the separation power, these four event shape
variables are often combined into a Fisher discriminant $\cal{F}$~\cite{fisher}. In some analyses,
neural network algorithm is used to combine information from a set of event shape variables, including
a set of energy flow cones. Finally, the Fisher/neural net variables are combined 
with kinematic variables in a maximum likelihood fit to determine 
simultaneously the signal yield and the charge asymmetry. 

\section{Measurement of the $CP$ Asymmetry}
In this section we review the results on a few selected measurements 
of $CP$ asymmetries performed
with the BaBar detector. Unless otherwise stated, all
measurements presented here are preliminary. 


\subsection{First Observation of Direct $CP$ Violation in B decays: $A_{K \pi}(B^{0} \to K^+ \pi^-)$$^{~8}$}
The decay $B^{0} \to K^{+} \pi^{-}$ occurs through two different diagram types of diagram (``penguin''
and ``tree''), which carry different weak phases and, in general, different strong phases.
The direct $CP$ violating asymmetry for this mode is defined by
\begin{equation}
A_{K \pi} = \frac{n_{K^{-}\pi^{+}} - n_{K^{+}\pi^{-}}}{n_{K^{-}\pi^{+}} + n_{K^{+}\pi^{-}}}
\label{eq:acpkpi}
\end{equation}
where $n_{K^{-}\pi^{+}}$ and $n_{K^{+}\pi^{-}}$ are the measured yields for the two final 
states. We require that each track has an associated Cherenkov-angle ($\theta_{c}$) 
measured with detector of internally reflected Cherenkov light (DIRC). This information
is used to separate kaons and pions in a maximum-likelihood fit that determines signal
and background yields corresponding to the four distinguishable final states ($\pi^{+}\pi^{-}$, 
$K^{+}\pi^{-}$, $K^{-}\pi^{+}$ and $K^{+}K^{-}$). The likelihood of any event is
obtained by summing the product of the ``event yield'' ($n_{K^{\pm} \pi^{\mp}}$) 
and the probability density function (PDF) which use
observables $m_{ES}$, $\Delta E$, $\cal{F}$, $\theta_{c}^{+}$, $\theta_{c}^{-}$. 
The $\theta_{c}^{\pm}$ PDFs are
obtained from a sample of approximately 430000 $D^{*+} \to D^{0} \pi^{+} ( D^{0} \to K^{-} \pi^+)$
decays reconstructed in data as shown in Fig.~1(a). The $K^{\mp}\pi^{\pm}$ yields are 
parametrized as $n_{K^{\pm} \pi^{\mp}} =  n_{k \pi}(1 \mp A_{K \pi})/2$, where
$n_{k \pi}$ is the total yield. Based on a data sample of 253 fb$^{-1}$, a fit to the signal events
measures $n_{K \pi}= 1606 \pm 51$, $A_{K \pi}= -0.133 \pm$0.030(stat)$\pm$0.009(syst)
and the background asymmetry $A_{K \pi}^{b} = 0.001 \pm 0.008$. 
As shown in Fig. 1(b,c), a clear enhancement of $K^{+} \pi^{-}$ (solid histogram) 
is observed in the distribution with $M_{es}> $ 5.27 GeV 
whereas the charge asymmetry is negligible for the background events with $m_{ES} <$5.27 GeV.
As part of the consistency check, we divided the entire data sample into the approximate periods in 
which the data were recorded. We find $A_{K \pi}<0$ (Fig.1(d)) and background asymmetries
consistent with zero in each data set.
\begin{figure}
\begin{center}
\epsfig{figure=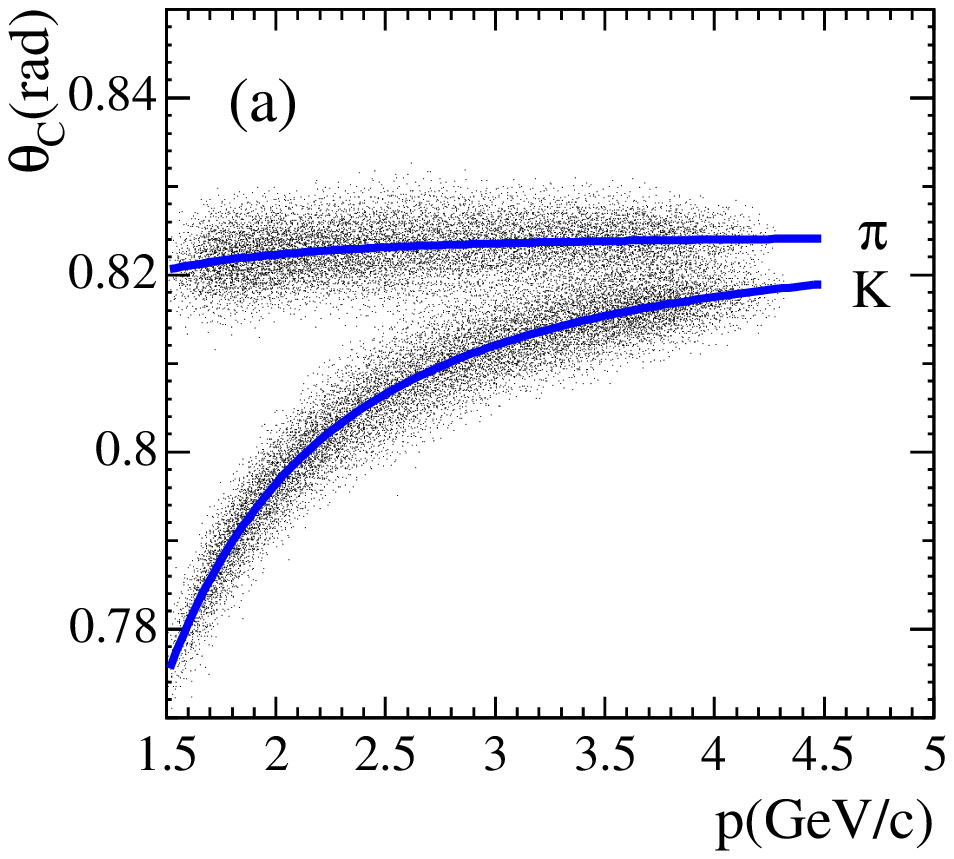,height=1.7in}
\epsfig{figure=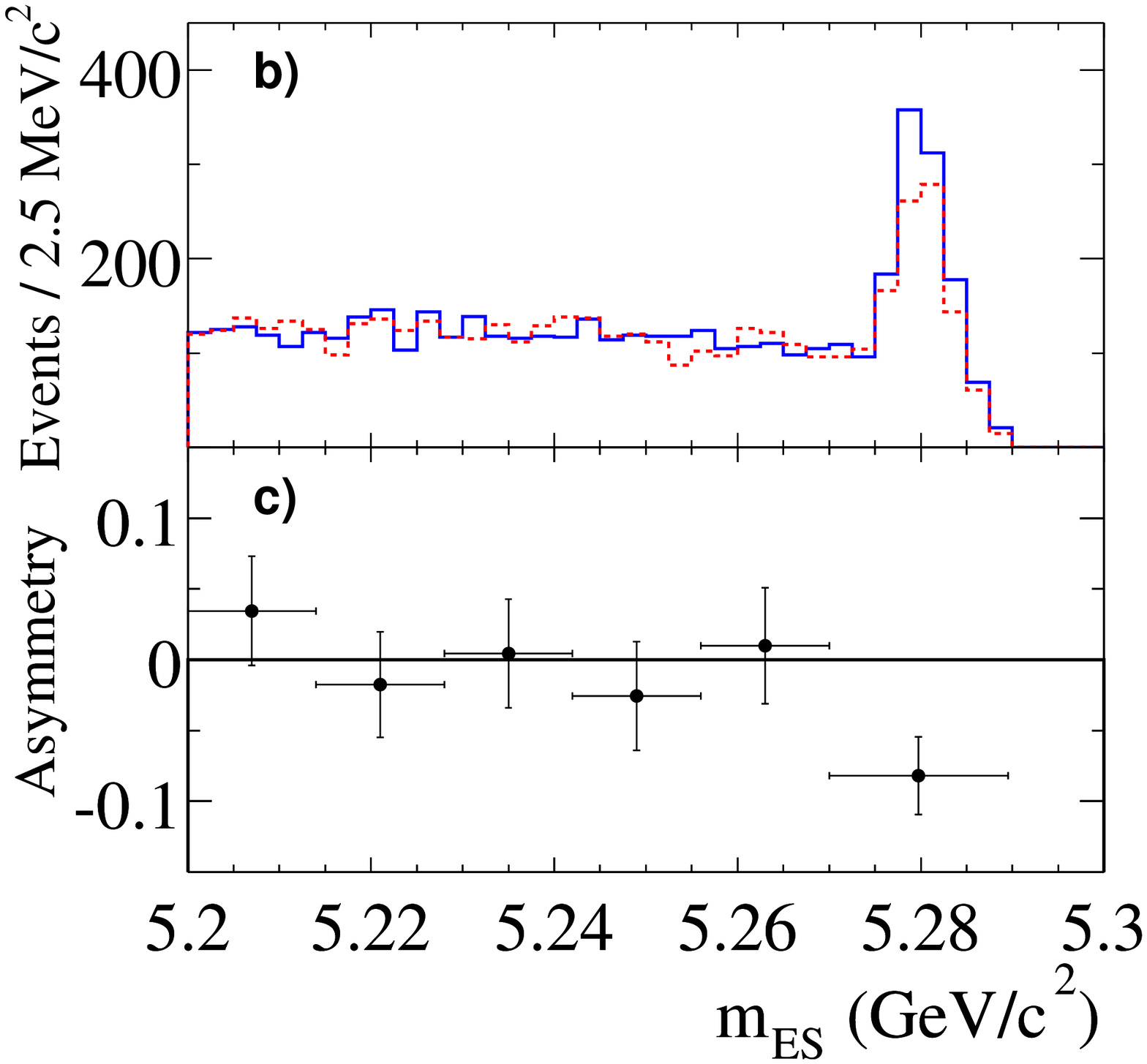,height=1.7in}
\epsfig{figure=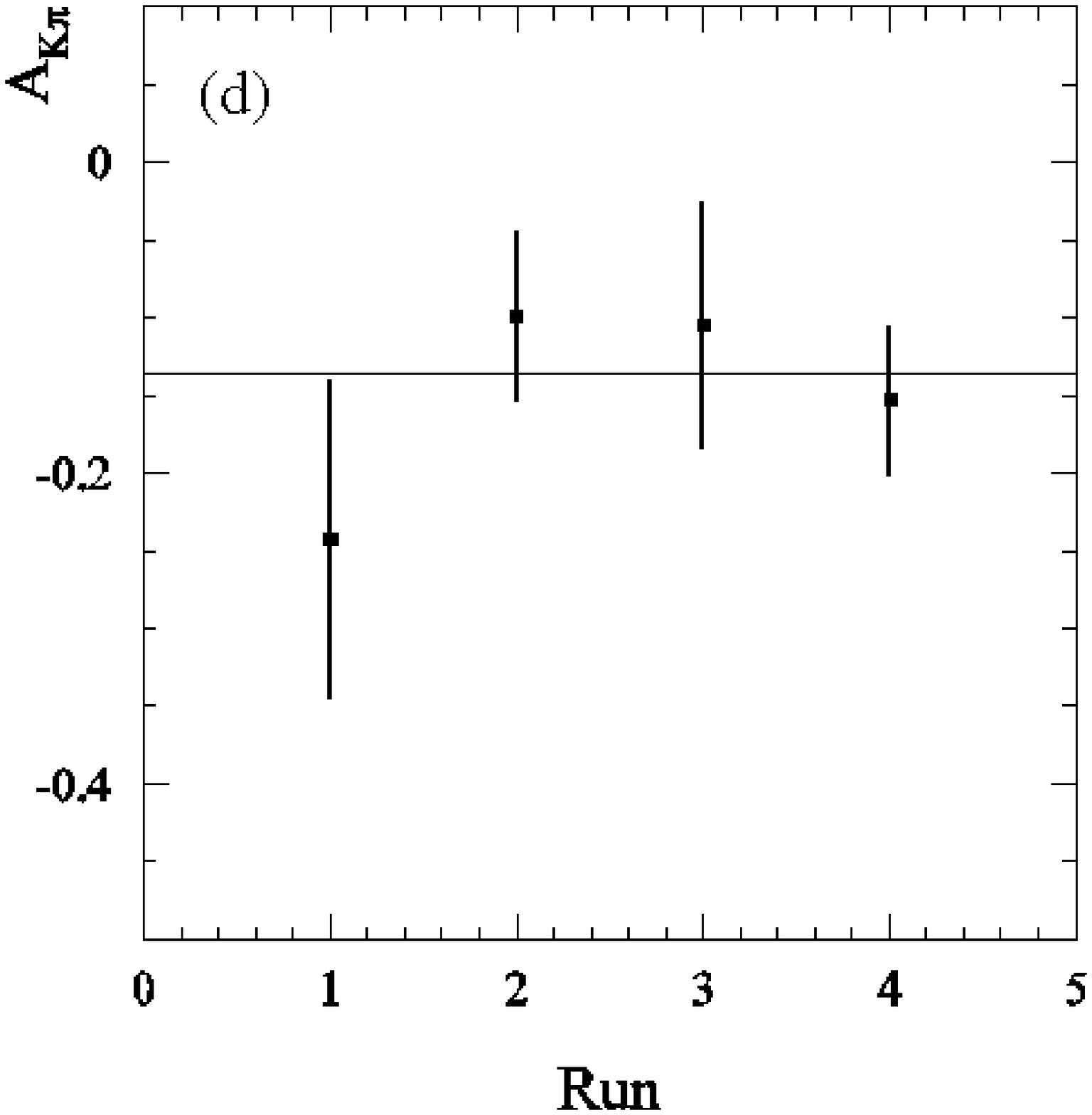,height=1.7in}
\end{center}
\caption{(a) The measured Cerenkov angle distribution for pions and kaons from 
$D^{*} \to D^{0} \pi$, and $D^{0} \to K \pi$ decays from data. The lines show the
expected angle $\theta_{c}$ as a function of laboratory momentum. (b) Distribution of
$M_{ES}$ enhanced in $K^{+} \pi^{-}$ (solid histogram) and $K^{-} \pi^{+}$ (dashed histogram).
(c) $CP$ asymmetry calculated for different ranges of $M_{ES}$. (d) Measured $CP$ Asymmetry
in different data samples.}
\end{figure}

Five years after direct $CP$ violation was observed in K-meson decays, this
measurement establishes direct $CP$ violation in $B^{0}$-meson system at the level of 4.2 
standard deviations.The Belle collaboration has recently reported 
an updated measurement~\cite{kpi_belle} of $A_{K \pi}= -0.101 \pm 0.025 \pm 0.005$ 
which confirms our observation. The results are consistent within the expected range
of SM.


\subsection{First Measurement of $A_{CP}(B^{+} \to K^+ K^{0}_{S} K^{0}_{S})$$^{~10}$}
Since this process is expected to be dominated by $b \to s \bar{s} s$ loop transition, 
the SM prediction of $A_{CP}$ is zero. Hence, this could be a place where one 
may observe a signal for new physics. The measurement of $A_{CP}(B^{+} \to K^+ K^{0}_{S} K^{0}_{S})$ is
based on 122 million $B \bar{B}$ pairs. An unbinned extended maximum likelihood fit  
was performed to the data sample where the event yields are split by the charge 
and extracted separately for signal, continuum and peaking $B$ background.
We have reconstructed a total of 6144 signal events in this mode and
measured value of $A_{CP}(B^{+} \to K^{+} K_{S}^{0} K_{S}^{0}) =  -0.04 \pm 0.11 \pm 0.02$.
The measured $CP$ asymmetry is consistent with SM prediction.


\subsection{Measurement of $A_{CP}(b \to s \gamma)$$^{~11}$}
In the SM the inclusive decay $b \to s \gamma$ is a flavor changing neutral
current process described by a radiative loop diagram and the predicted direct $CP$
asymmetry is close to zero~\cite{hurth}. Direct $CP$ asymmetry is calculated from
\begin{equation}
A_{CP} = \frac{1}{\langle D \rangle}(\frac{(n - \bar{n})}{(n + \bar{n})} - \frac{\Delta D}{2}) - A_{CP}^{det}
\label{eq:acpkksks}
\end{equation}
 where $n$ and $\bar{n}$ are the numbers of observed $b \to s \gamma$ and $\bar{b} \to \bar{s} \gamma$
events after peaking background is subtracted, $\Delta D$ is the difference in the wrong flavor-fraction
between $b$ and $\bar{b}$ decays, and $\langle D \rangle$ is the dilution factor from the average wrong flavor-fraction.
The correction term $A_{CP}^{det}$ is the flavor asymmetry in the detector
and measured to be $-0.014 \pm 0.015$. Signal events are reconstructed as the sum of eight exclusive
final states: $B^{-} \to K^{-} \pi^{0} \gamma$, $K^{-} \pi^{+} \pi^{-} \gamma$, $K^{-} \pi^{0} \pi^{0} \gamma$,
$K^{-} \pi^{+} \pi^{-} \pi^{0} \gamma$ and $B^{-} \to K_{S}^{0} \pi^{-} \gamma$, $K_{S}^{0} \pi^{-} \pi^{0} \gamma$,
$K_{S}^{0} \pi^{-} \pi^{0} \pi^{0} \gamma$ and $K_{S}^{0} \pi^{-} \pi^{+} \pi^{-} \gamma$.

Based on a measurement with 89 million $B \bar{B}$ pairs, our recently published results are 
$A_{CP}(b \to s \gamma) = 0.025 \pm 0.050 \pm 0.015 $  for the total sample and
$A_{CP}(b \to s \gamma) = -0.04 \pm 0.10 \pm 0.02$ for the lepton-tagged sample. This
value is consistent with SM prediction.


\subsection{Measurement of $A_{CP}(B \to K^{*} \gamma)$$^{~13}$}
Unlike inclusive decays, exclusive $B \to K^{*} \gamma$ decay rates have large uncertainties
due to non-perturbative hadronic effects, limiting their usefulness for probing new
physics. However, the interest in measuring $A_{CP}(B \to K^{*} \gamma)$ clearly lies in 
making a stringent test of the SM which predicts this value to be less than 1\%.
We reconstruct $B^{0} \to K^{*0} \gamma$ in the $K^{*0} \to K^{+} \pi^{-}$ mode
and $B^{+} \to K^{*+} \gamma$ in the $K^{*+} \to K^{+} \pi^{0}, K_{S}^{0} \pi^{+}$ modes.
Using a sample of 88 million $B \bar{B}$ events, we measure a combined direct
$CP$ asymmetry of $-0.013 \pm 0.036 \pm 0.010$ which is, within experimental error, 
consistent with SM prediction.

\section{Conclusion and Prospects}
We reported the first observation of direct $CP$ violation in the $B$ meson 
decay in $B^0 \to K^{-} \pi^{+}$.
We do not observe any significant $CP$ asymmetry in other decay modes such as 
$B^{+} \to K^{+} K_{S}^{0} K_{S}^{0}$, 
$b \to s \gamma$ and $B \to K^{*} \gamma$. Improved statistics will  
reveal new insights in the ongoing searches for direct $CP$ violation. With the
excellent running performances of the BaBar detector and lot more luminosity from PEP-II to come, 
we look forward to an exciting time that will confirm or refute the
SM and its description of $CP$ violation.

\end{document}